# Whack-A-Mole Model: Towards a Unified Description of Biological Effects Caused by Radiation Exposure

Yuichiro Manabe[1],

Takahiro Wada[2],

Yuichi Tsunoyama[3],

Hiroo Nakajima[4],

Issei Nakamura[5],

Masako Bando[6,7]

[1]Division of Sustainable Energy and Environmental Engineering, Graduate School of Engineering, Osaka University, Suita, Osaka 565-0871, Japan

[2]Department of Pure and Applied Physics, Faculty of Engineering Science, Kansai University, Suita, Osaka 564-8680, Japan

[3]Division of Biology, Radioisotope Research Center, Kyoto University, Sakyo, Kyoto 606-8502, Japan

[4]Department of Radiation Biology and Medical Genetics, Graduate School of Medicine, Osaka University, Suita, Osaka 565-0871, Japan

[5]State Key Laboratory of Polymer Physics and Chemistry, Changchun Institute of Applied Chemistry, Chinese Academy of Sciences, Changchun, Jilin 130022, P. R. China

[6]Yukawa Institute for Theoretical Physics, Kyoto University, Kyoto 606-8502, Japan

[7]Research Center for Nuclear Physics (RCNP), Osaka University, Ibaraki, Osaka 567-0047, Japan

Abstract

  We present a novel model to for estimating biological effects caused by artificial radiation exposure, i.e., the Whack-A-Mole (WAM) model. It is important to take into account the recovery effects during the time course of cellular reactions. The inclusion of dose-rate dependence is essential in the risk estimation of low-dose radiation, while nearly all the existing theoretical models rely on the total dose dependence only. By analyzing experimental data of the relationship between the radiation dose and the induced mutation frequency of five organisms, namely, mouse, *Drosophila*, chrysanthemum, maize, *Tradescantia*, we found that all the data can be reproduced by the WAM model. Most remarkably, a scaling function, which is derived from the WAM model, consistently accounts for the observed mutation frequencies of the five organisms. This is the first rationale to account for the dose rate dependence as well as to provide a unified understanding of a general feature of organisms.

# 1. Introduction

The discovery of mutation induced by artificial irradiation was first made by Muller in 1927 while investigating the genetic effects of X-rays on *Drosophila*[1]. Almost at the same time, in 1928, Stadler found that artificial mutation occurs in plants as well, through the study of mutation in maize and barley caused by X-ray treatments[2]. Until then, it had been considered that mutations, which trigger biological evolution, occur only naturally. The experimental support of artificial mutation accelerated the development of the theory of evolution and molecular biology; vast amounts of studies have emerged on the subject of radiation exposure and genetic mutation. In particular, following the work of Muller, experimental data on *Drosophila* accumulated and confirmed the concept of the "linear no threshold (LNT) " hypothesis, which states that the mutation caused by ionizing radiation increases in proportion to total dose only, independently of dose rate. This observation had a strong impact not only on the scientific community but also on society; the LNT hypothesis has long been adopted as a scientific basis of international organizations aiming for radiation protection.

On the other hand, W. Russell of Oak Ridge National Laboratory proposed to test the validity of the LNT hypothesis in mice[3]. This is called the "mega-mouse" experiment, which investigated the frequency of transmitted specific seven locus mutations induced in mouse spermatogonial stem cells. Their results indicated that mutation frequency varies with the dose rate of the ionizing radiation even if the total dose is the same[4]. Moreover considerable data on plants also existed, indicating the dose rate effect[5, 6, 7]. These results contradict the LNT hypothesis. This is important if we want to compare the mutation frequencies of a variety of species, since human risk estimates of radiation at that time had been almost exclusively based on *Drosophila* studies until the mega-mouse experiments. Studies on mice have helped answer basic questions in radiation genetics and have given a clearer understanding of the genetic hazard of radiation exposure. However, although many theoretical and experimental studies have been performed, it has not yet been clarified whether the dose rate dependence exists and how we can understand two different results, namely, the *Drosophila* data of Muller and the mouse data of Russell.

Now, we are addressing two fundamental questions; the first is whether animals and plants have a common mechanism of reacting to radiation exposure, and the second is whether a quantitative description can be achieved for the dose rate dependence. In order to answer the above two questions, it is necessary to construct a realistic mathematical model that will take into account all the physical and biological procedures, such as stimulus-response physical reactions as well as the biological mechanisms occurring in organisms, such as cell proliferation and cell death. While nearly all existing theoretical models rely on the total dose dependence only, we would like to stress that the dependence of the dose rate on mutation frequency is the key ingredient in constructing a model for biological effects, because the dose rate dependence accounts for recovery effects during the time course of cellular reactions, and the number of mutated cells is determined by the time-dependent procedures of antagonism between DNA damage and repair[8, 9, 10].

In this paper, we present a mathematical model for estimating biological effects caused by artificial irradiation exposure, which was developed in previous papers[11, 12, 13]. We call this the "Whack-A-Mole" (WAM) model, which enables us to numerically calculate the mutation frequency of organisms. The explicit time dependence is essential for accommodating the strategy of survival mechanism, which prohibits the increase in the number of mutated cells. This markedly changes the concept of risk estimation of low-dose radiation.

Most remarkably, the scaling function, which is derived from the WAM model, accounts for the observed mutation frequencies of a variety of organisms in a unified way. The theoretical predictions are manifested by a self-explanatory figure (See Fig. 4), which is the first rationale to account for the dose rate dependence. Furthermore, it gives a unified understanding of a general feature of antagonism between destruction and reconstruction, which are commonly observed in various of organisms.

In Sect. 2, we explain the target theory and linear-quadratic model (LQM) of its developed version, and introduce five experimental data that we are going to analyze. In Sect. 3, we propose a general formalism of the WAM model and explain the characteristics of its solution. In sect. 4, we apply the present experimental data and show the numerical results. Section 5 is devoted to summary and discussion.

## 2. Historical Review of Theory and Experiment

On the theoretical side, the target theory, originally formulated by Lea, has been the basis of radiobiology. His textbook is still considered the bible in this field[8]. According to his theory, an individual quantum of radiation is absorbed at sensitive points (targets), and then normal cells change to mutated cells. Then, the number of normal cells $N_n$ is given by the following equation,

$$\frac{dN_n}{N_n} = -\frac{dD}{D_0}, \quad (1)$$

where $D$ and $D_0$ are the total dose of irradiation and the unit dose to produce one active event, respectively. Then, the number of mutated cells, $N_m$, is

$$N_m(D) = N_n(0) - N_n(0)e^{-\frac{D}{D_0}} + N_m(0) = N_n(0)(1 - e^{-\frac{D}{D_0}}) + N_m(0), \quad (2)$$

where we have added the initial condition at $D=0$, $N_m(0)$. We could further add a quadratic term, taking into account the two-hit contribution in addition to the one-hit contributions. This is actually because the simple LNT hypothesis was not able to reproduce the experimental data of a high total dose region. Thus, such quadratic term might be interpreted as a higher order correction term. Also, we could take into account a recovery system to reduce the number of mutated cells:

$$N_m(D) = \sigma N_n(0)\left[1 - e^{-(\frac{D}{D_0}+\frac{D^2}{D_1})}\right] + N_m(0) \rightarrow E(D) \equiv \frac{N_m(D)}{N_n(0)} = \sigma\left[1 - e^{-(\frac{D}{D_0}+\frac{D^2}{D_1})}\right] + E(0), \quad (3)$$

where $\sigma$ is the proportion of mutated cells that are not repaired, and the quadratic term $D^2/D_1$ is added to the exponent to account for the observed deviation from the classical target one-hit theory at high dose rates[9]. This equation is called the LQM, which was developed by Chadwick and Leenhouts, accounting for the recovery effects and so forth. In the low-dose region, the above formula is converted into its approximate form[14],

$$N_m(D) \approx N_n(0)(1 + \frac{D}{D_0} + \frac{D^2}{D_1}) + N_m(0). \quad (4)$$

This approximate form of Eq. (4) is also sometimes called LQM.

It is interesting to note that those models are considered to be applicable to a variety of processes, such as chromosomal aberrations, mutations, and carcinogenesis. Therefore, we can extend the notion of mutation frequency to more general cases denoted by the "effect (E)" that is caused by outer stimuli,

$$E(D) \approx E(0) + aD + bD^2 . (5)$$

Note that, in the above formula, the dependence of dose rate $d \equiv dD/dt$ on the effect $E$ was still not considered despite the observation of dose rate dependence in mega-mouse experiments by Russell and Kelly[4]. This may be because of the strong notion embraced through a long history. Later, we shall compare this to our model, which will help in understanding the essential difference between LQM and the WAM model.

As for the experimental situation, we found few data of mutation frequencies, which addressed the dose rate dependence; most experimentalists seem to have been interested only in the total dose dependence. This would be due to the LNT stereotype dogma. In Fig. 1, we summarize the data on mutation frequencies of five species, namely, mouse, *Drosophila*, maize, *Tradescantia*, and chrysanthemum[4, 5, 6, 7, 15, 16, 17]. The data points of each species are scattered in the $(D, F)$ plane, with $D$ and $F$ being the total amount of dose (horizontal axis) and mutation frequency (vertical axis), respectively. We find that the data can be classified into three groups; *Tradescantia* takes values on the order of $10^{-2} - 10^{-1}$, while those of mouse are on the order of $10^{-5} - 10^{-4}$, with those of the others being located in between. Plants are mutated more easily than animals, and spermatogenetic cells might have more recovery capability (mouse) than sperm cells.

When we inspect the data of each species more closely, we find that there are some data points that take different in the case of mutation frequencies even though the total dose is the same. For example, mutation frequencies for a total dose of 3 Gy, there are two data points, namely, exposure with dose rate of 54 Gy/hr leads to a fivefold higher mutation frequency than that with a dose rate of 0.00257 Gy/hr; a higher dose rate yields a higher mutation frequency even if the total dose is the same. In the next section, we shall see that the values predicted with the WAM model can reproduce all the data for different dose rates in a unified way.

Fig. 1

### 3. Whack-A-Mole Model

Let us consider a system, i.e., a tissue or an organ, consisting of normal and mutated cells, the numbers of which are denoted as $N_n$ and $N_m$ respectively. In general, a system is full of cells whose density is roughly $\rho = 10^{11} / \text{kg}$. Thus, the maximum number of cells in a system with its mass $M$ is $N_{max} = \rho M$.

Let the theory for radiation exposure be described in terms of the kinetic reaction of irradiated DNA molecules. Its reaction mechanisms involve a broad range of time scales from nanoseconds to years. In addition, and we are often encounter complicated problems because of the diversity of living organisms. In any case, however, the first stage of the reaction pathway of the breaking of DNA molecules in picoseconds is mainly caused by free hydroxyl radicals due to ionizing radiation, followed by a longer-time-scale process of DNA mutation in cell cycles, during which biological mechanisms such as cell proliferation and cell death are involved[18]. Thus, the key ingredient is the time dependence of $N_m$, because the damage caused by the irradiation and environmental stimuli to DNA molecules is strongly reduced in reference to counteracting effects. This controls the reaction of the system in accordance with the dose rate without invoking the total dose $D$. In general, they are described accordingly by the differential equations

$$\frac{dN_n}{dt} = R_{nn}N_n + R_{nm}N_m,$$

$$\frac{dN_m}{dt} = R_{mn}N_n + R_{mm}N_m, \quad (6)$$

where the matrix $R$ represents reaction rate, $R_{nn}$ corresponds to the proliferation or cell death of normal cells, $R_{nm}$ denotes the rate of repair from mutated cells to normal ones, $R_{mn}$ indicates the mutation of normal cells, and $R_{mm}$ corresponds to the proliferation or cell death of mutated cells. Note that $R_{nm} = 0$ because of the definition of mutation, namely, mutations result from unrepaired damage to DNA or to RNA genomes. Indeed, the process from normal to mutated cells might have several steps, and if we introduce damaged cells as an intermediate stage, we would need to include the repair mechanism itself. However, these steps may be described by many parameters of the reaction rates, and they may depend on complex mechanism and the surrounding conditions. Here, we do not specify all the biological reactions because we do not wish to include a large number of rate constants, which actually cannot be determined or have a large uncertainty. Instead, we only focus on the average effective rate of DNA mutation due to all these relevant reactions.

Here, we compare our model with experiments on mutation frequency under the condition that, before irradiation, the system already has a few mutated cells caused by natural surrounding stimulus (which is called "spontaneous mutation"). Let us restrict ourselves to the case of low-dose rate and to the system dominated by normal cells, namely, $N_n(t=0) \approx N_{\max}$, before irradiation starts. The mutation frequency is defined as the ratio of $N_m$ to $N_n(t=0)$, which is on the order of at least $10^{-2}$ at $t=0$; $N_n(t=0) \approx N_{max}$, $F(0) \equiv \dfrac{N_m(t=0)}{N_{max}} \ll 1$. Thus, we are led to the following simple form,

$$F(t) \equiv \frac{N_m(t)}{N_n(t=0)} \approx \frac{N_m(t)}{N_{max}}, \quad (7)$$

Then, $N_m(t)$ in Eq. (6) is decoupled, yielding the following expression,

$$\frac{dN_m(t)}{dt} = R_{mn} N_{max} + R_{mm} N_m(t) \rightarrow \frac{dF(t)}{dt} = R_{mn} + R_{mm} F(t). \quad (8)$$

Now, we start an artificial irradiation with its dose rate $d(t)$. Then, we can express the effects of the stimulus-response procedure as

$$R_{mn} = a_0 + a_1 d(t), \; R_{mm} = -(b_0 + b_1 d(t)) \rightarrow \frac{dF(t)}{dt} = [a_0 + a_1 d(t)] - [b_0 + b_1 d(t)] F(t), \quad (9)$$

where $a_0$, $a_1$, $b_0$ and $b_1$ are parameters representing the characteristics of species; the first term represents the increase in mutation and the second term represents the decrease due to cell death such as a type of programmed cell death (apoptosis). We name our model the WAM Model, which takes into account a striking function of organisms to kill damaged cells like the whacking-a-mole game, where the increase in the number of mutated cells, coming from the first terms with reaction rate $R_{mn} \equiv a_0 + a_1 d(t)$ is cancelled by the second term with the rate $R_{mm} \equiv -(b_0 + b_1 d(t))$. This indicates the existence of an upper limit of the mutation frequency; the increase in mutation frequency will stop with the stationary condition

$$F_{stationary}(t) = \frac{a_0 + a_1 d(t)}{b_0 + b_1 d(t)}. (10)$$

If we have no artificial stimulus, the above mutation frequency reproduces the so-called "spontaneous mutation frequency",

$$R_{mn} = a_0, \ R_{mm} = -b_0 \ \rightarrow F(\infty, d=0) \equiv F_s = \frac{a_0}{b_0}. \quad (11)$$

There are only four parameters, which are for the moment to be determined from the observed data.

Here, let us consider the case where the artificial constant irradiation with its dose rate $d$ starts at $t=0$ as shown in Fig. 2; then, the solution of Eq. (8) is analytically obtained,

$$F(t) = F(\infty)\left(1 - e^{-(b_0 + b_1 d)t}\right) + F(0)e^{-(b_0 + b_1 d)t}, \quad \text{for } 0 \leq t, \quad (12)$$

$$F(0) \equiv F(t=0), \ F(\infty) \equiv \frac{a_0 + a_1 d}{b_0 + b_1 d}.$$

This function is of similar form to one of those known as "growth functions", which are often used for empirically fitting plant growth data[19]. The form of Eq. (12) corresponds to monomolecular growth function, not logistic function. The growth function has some upper bound over time. The only difference of the Eq. (12) from the growth function is that the dose-rate-dependent terms are included in the exponent $\tau \equiv (b_0 + b_1 d)t$ and the upper bound value $F(\infty)$. The parameters $a_0$ and $a_1$ represent the result of a stimulus-response procedure, the radio sensitivity for changing normal to mutated cells for the former, and that for the death of mutated cells caused by irradiation for the latter.

Fig. 2

Let us here consider the special case where the initial condition is controlled by the value of spontaneous mutation,

$$F(0) = F_s = \frac{a_0}{b_0} \to F(t) = \frac{a_0 + a_1 d}{b_0 + b_1 d}\left(1 - e^{-(b_0 + b_1 d)t}\right) + \frac{a_0}{b_0} e^{-(b_0 + b_1 d)t}, \quad \text{for } 0 \le t, \quad (13)$$

which represents the situation usually assumed by experimentalists. The exception is, for example, the case where the object already has many mutated cells after a certain treatment, or the object does not reach the stationary state and still the number of mutated cells is increasing spontaneously. If one writes Eq. (13) in terms of the dimensionless time $\tau \equiv (b_0 + b_1 d)t$ and the upper bound $F(\infty) \equiv \frac{a_0 + a_1 d}{b_0 + b_1 d}$, Eq. (13) can be expressed as

$$F(t) = F(\infty)\left(1 - e^{-\tau}\right) + F_s e^{-\tau}, \quad \text{for } 0 \le \tau, \quad (14)$$

where we use the control value $F_s$. Here, we define the dimensionless time $\tau$, which provides us with the index for estimating critical time.

In the region $\tau = (b_0 + b_1 d)t \ll 1$, we have from Eq. (13),

$$\begin{aligned}
\lim_{(b_0 + b_1 d)t \to 0} F(t) &= \frac{a_0 + a_1 d}{b_0 + b_1 d}\left(1 - e^{-(b_0 + b_1 d)t}\right) + \frac{a_0}{b_0} e^{-(b_0 + b_1 d)t} \\
&= \frac{a_0 + a_1 d}{b_0 + b_1 d}\left(1 - (1 - (b_0 + b_1 d)t)\right) + \frac{a_0}{b_0}\left(1 - (b_0 + b_1 d)t\right) \\
&= (a_0 + a_1 d)t + \frac{a_0}{b_0}(1 - (b_0 + b_1 d)t) = \left(a_1 - \frac{a_0}{b_0} b_1\right) d \cdot t + \frac{a_0}{b_0}, \\
&\text{for } \tau = (b_0 + b_1 d)t \ll 1, \quad (15)
\end{aligned}$$

where the dose rate dependence appears only in the first term associated with time $t$.

The total dose $D = d \cdot t$ then (13) yields

$$F(D) = \left(a_1 - b_1 F_s\right)D + F_s \quad \text{for } \tau = \frac{D(b_0 + b_1 d)}{d} \ll 1. \quad (16)$$

Thus, under such situation, the mutation frequency depends only on the total dose, $D$ instead of $d \cdot t$. Then, the slope with respect to $D$ is constant, independent of the dose rate $d$. This is usually what happens in a low-dose limit. In this sense, the LNT hypothesis is realized in the low-dose limit. As for the slope, it is determined not only by the radio sensitivity but also by the repairing contribution term $-b_1 F_s$, which is proportional to the number of mutated cells. The slope is just a result of a counterbalance between the increase and decrease in the number of mutated cells. However, it should be mentioned that the dimensionless time $\tau = (b_0 + b_1 d)t$ depends on the dose rate, and above $\tau = 1$, the linear approximation form is no longer valid and the linearity collapses very quickly. In particular, it quickly deviates from the linear extrapolated prediction if the dose rate $d$ is extremely small.

When $t \to \infty$, $F(t)$ approaches the stationary value, which is already seen in Eq. (12),

$$\lim_{t \to \infty} F(t) \equiv F(\infty) = \frac{a_0 + a_1 d}{b_0 + b_1 d}, \quad (17)$$

which depends on the dose rate. If one combines the above approximate forms, we can roughly express WAM model in terms of a two-component model (2CM), i.e.,

$$F_{2cm}(t) \equiv \begin{cases} \left(a_1 - b_1 F_s\right)dt + F_s. & \text{for } \tau = (b_0 + b_1 d)t \leq 1 \\ \dfrac{a_0 + a_1 d}{b_0 + b_1 d}, & \text{for } 1 \leq \tau = (b_0 + b_1 d)t \end{cases}. \quad (18)$$

To guide the eye, we illustrate a typical figure of 2CM by comparing an exact solution of $F$ from which a global structure of this model can be grasped. Here, we use the parameters obtained from the data of mouse, which has some structural resemblance to a human being, so the parameter set would give a concrete impression by comparing the so-called LNT or LQM. Experimental data are taken from Russell[4].

Fig.3

The black points are for $d \sim 10^{-3}$ [Gy/hr] and the red points are for $d \sim 10^{1}$ [Gy/hr]. The colors of the solid lines correspond to the prediction of our WAM model for $d = 5.0 \times 10^{+1}$ (red) and $5.0 \times 10^{-3}$ [Gy/hr] (black). Here, we have used the concrete values of the parameters $a_0$, $a_1$, $b_0$ and $b_1$ of Eq. (13) are shown in Table I. The dotted black and red lines denote the corresponding 2CM results of Eq. (18). For comparison, we also add the blue solid curve, which represents the LQM fit using Eq. (5), together with the dotted blue line, i.e., linear line with its slope $a$, which is just the approximate linear form of the WAM model shown by Eq. (15) in the low-dose limit. The BEIR VII report indicates that this can be identified as the linear line of mutation frequency for the data of chronic exposure, while that for the acute exposure in the form $E(D) \approx E(0) + aD$, the red solid curve corresponds to the LQM prediction for acute exposure.

The essential feature of the WAM model is that mutation frequency does not increase indefinitely but it approaches a stationary value after a critical time (or critical total dose),

$$\tau_c = 1 \rightarrow t_c = (b_0 + b_1 d)^{-1}, \quad D_c = dt_c = \frac{d}{b_0 + b_1 d}. \quad (19)$$

In Fig. 3, the flat part of 2CM changes its maximal value depending on the dose rate; the higher the dose rate, the higher the maximal value. Thus the prediction of the LNT hypothesis deviates quite markedly from the experimental points. In this sense, the mutation frequency does not increase even if the accumulated time or accumulated total dose increase in the case of low-dose-rate irradiation.

## 4. Scaling Results

In order to compare the WAM model predictions with experimental data, it is convenient to cast Eq. (12) into

$$\Phi(\tau) \equiv \frac{F(t) - F(0)}{F(\infty) - F(0)} (1 - e^{-\tau}), \quad \tau \equiv (b_0 + b_1 d) t,$$

$$\lim_{t \to \infty} F(t) \equiv F(\infty) = \frac{a_0 + a_1 d}{b_0 + b_1 d}, \quad (20)$$

where $\Phi(\tau)$ is the renormalized mutation frequency in terms of the dimensionless time $\tau$. This indicates that, in general, mutation frequencies consist of the universal scaling function $\Phi(\tau)$.

Now that we have the scaling function $\Phi(\tau)$, let us apply our WAM model to the data shown in Fig. 1. The procedure is as follows:

1) Fix the four parameters $a_0$, $a_1$, $b_0$ and $b_1$ using the data of each species by $x^2$ fitting.
2) Calculate $\Phi(\tau)$ by inserting those parameters into F and $\tau$ in Eq. (12).
3) Plot all the data points in $(\tau, \Phi)$ plane.

The parameters determined from the data of five species are summarized in Table I.

Table I.

Fig. 4.

The results are shown in Fig. 4, with the theoretical curve that illustrates how our scaling function reproduces all the experimental data. It indicates that the WAM model predictions are in good agreement with the experimental data without classifying the dose rate. The experimental data of mutation frequencies fall on the predicted curve of the universal scaling function $\Phi(\tau)$.

## 5. Summary and Discussion

We have introduced the dimensionless time and the scaling function of the renormalized mutation frequency. Remarkably, our fitting was performed irrespective of the diversity of species, ranging from animals to plants. We have shown that the observed dependence of the dose rate on DNA mutation frequencies of the five species cannot be explained by classical theories.

Thus, our theory suggests that dose rate is a fundamental measure for studying irradiation as deduced from the systematic relationship between the dose rate and the induced mutation frequency. Vilenchik and Knudson showed the relationship between the dose rate and the mutation frequency in the mega-mouse project[20]. However, their method is polynomial regression, which is essentially different from ours.

We can extend our model to other species including viruses as well as human beings. Also, we can apply the WAM model to a variety of processes, such as chromosomal aberrations, mutations, and carcinogenesis, since antagonism between DNA damage and repair is commonly seen in various living organisms.

We hope that the WAM model opens the door towards a unified understanding of mutation frequencies across all species. We believe that our work will lead to make quantitative estimation of biological effects caused by artificial irradiation in a unified way, and lead us to revisit the LNT hypothesis for DNA mutation induced by artificial irradiation.

**Acknowledgements**

This work was supported by Toyota Physical and Chemical Research Institute Scholars.

Table I. Four parameters determined from the data.

Fig. 1. Summary of the five experimental data: F(D) versus the total dose D. (Color)

Fig. 2. Time schedule of irradiation

Fig. 3. Comparison of the values estimated from the approximate two-component model with those of the WAM model, together with those obtained using the LQM. The black points are for $d \sim 10^{-3}$ [Gy/hr] and the red points are for $d \sim 10^{1}$ [Gy/hr]. The solid black line is the WAM model predictions for $d = 5.0 \times 10^{-3}$ [Gy/hr] and the solid red line is for $d = 5.0 \times 10^{1}$ [Gy/hr]. The dotted black and red lines denote the corresponding 2CM results. The blue solid curve represents the LQM fit using Eq. (5) and the dotted blue line denotes $E(D) \approx E(0) + aD$. (Color)

Fig. 4. Comparison of experimental data with the theoretical curve in $[\tau, \Phi(\tau)]$ plane. (Color)

**Table 1. The set of 4 parameters determined from the data**

|  | Mouse | Drosophila | Maize | Chrysanth-emum | Tradescan-tia |
|---|---|---|---|---|---|
| $a_0$ [1/hour] | $3.2 \times 10^{-8}$ | $3.5 \times 10^{-5}$ | – | – | $2.9 \times 10^{-2}$ |
| $a_1$ [1/Gy] | $3.0 \times 10^{-5}$ | $2.0 \times 10^{-3}$ | $2.0 \times 10^{-3}$ | $6.5 \times 10^{-3}$ | $1.6 \times 10^{-1}$ |
| $b_0$ [1/hour] | $1.0 \times 10^{-3}$ | $1.4 \times 10^{-2}$ | $1.8 \times 10^{-1}$ | $4.5 \times 10^{-3}$ | $6.9 \times 10^{-1}$ |
| $b_1$ [1/Gy] | $1.4 \times 10^{-1}$ | $1.0 \times 10^{-4}$ | – | – | $1.6 \times 10^{-1}$ |

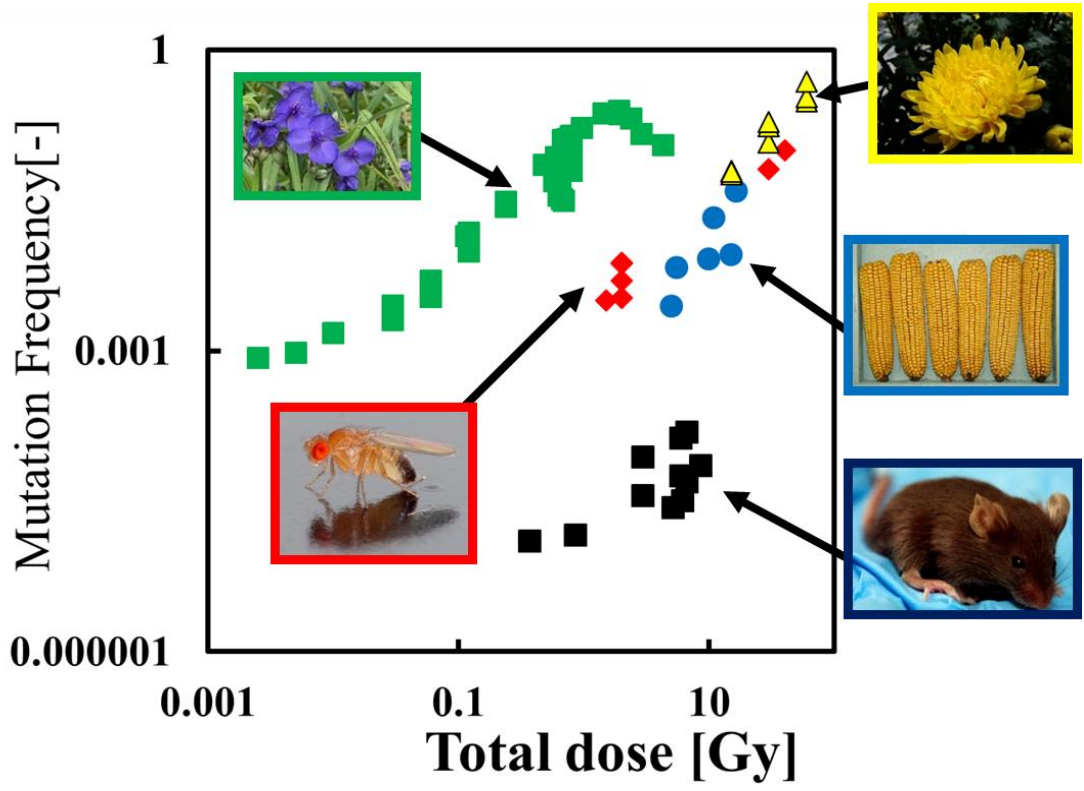

Fig. 1. Summary of the 5 experimental data F(D) versus the total dose D.

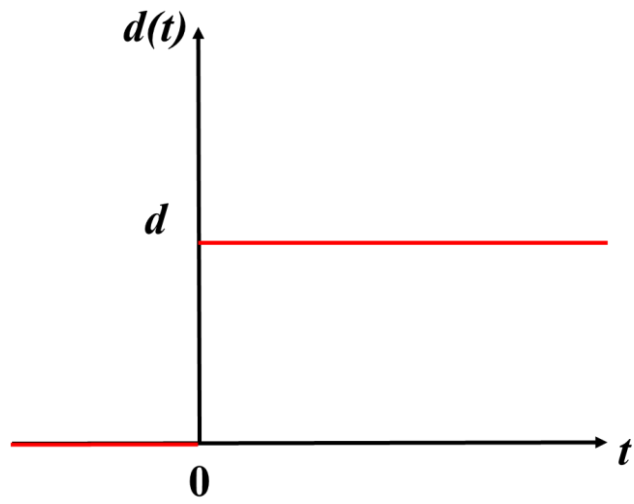

**Fig. 2. Time schedule of irradiation**

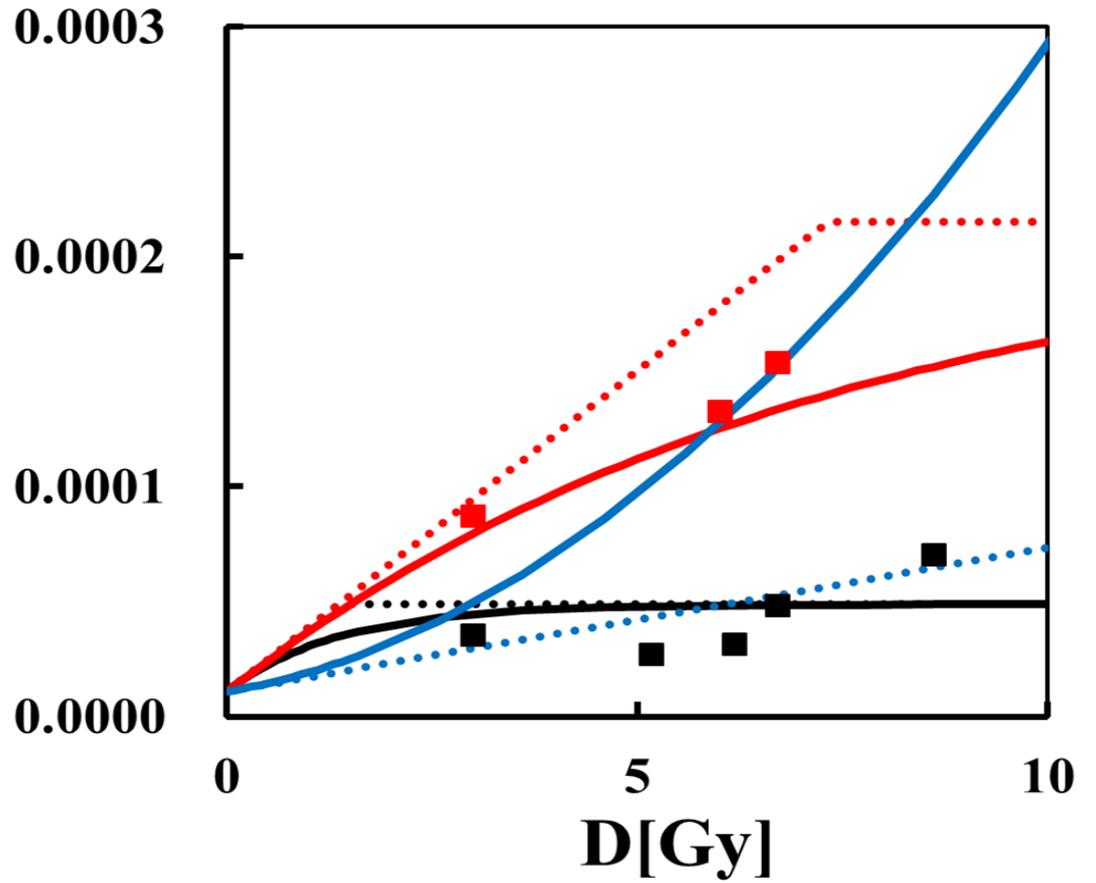

**Figure 3　Comparison of the vales estimated from approximate two component model with the one of exact ones, together with the one obtained by LQ model. The black points are for $d \sim 10^{-3}$ [Gy/hr] and red points, $d \sim 10^{1}$ [Gy/hr]. The solid black line is WAM predictions for $d = 5.0 \times 10^{-3}$ [Gy/hr] and the solid red line, $d = 5.0 \times 10^{1}$ [Gy/hr]. The dotted black and red line denote the corresponding 2CM results. The blue solid curve represents LQM fit using Eq. (5) and the dotted blue line, $E(D) \approx E(0) + aD$.**

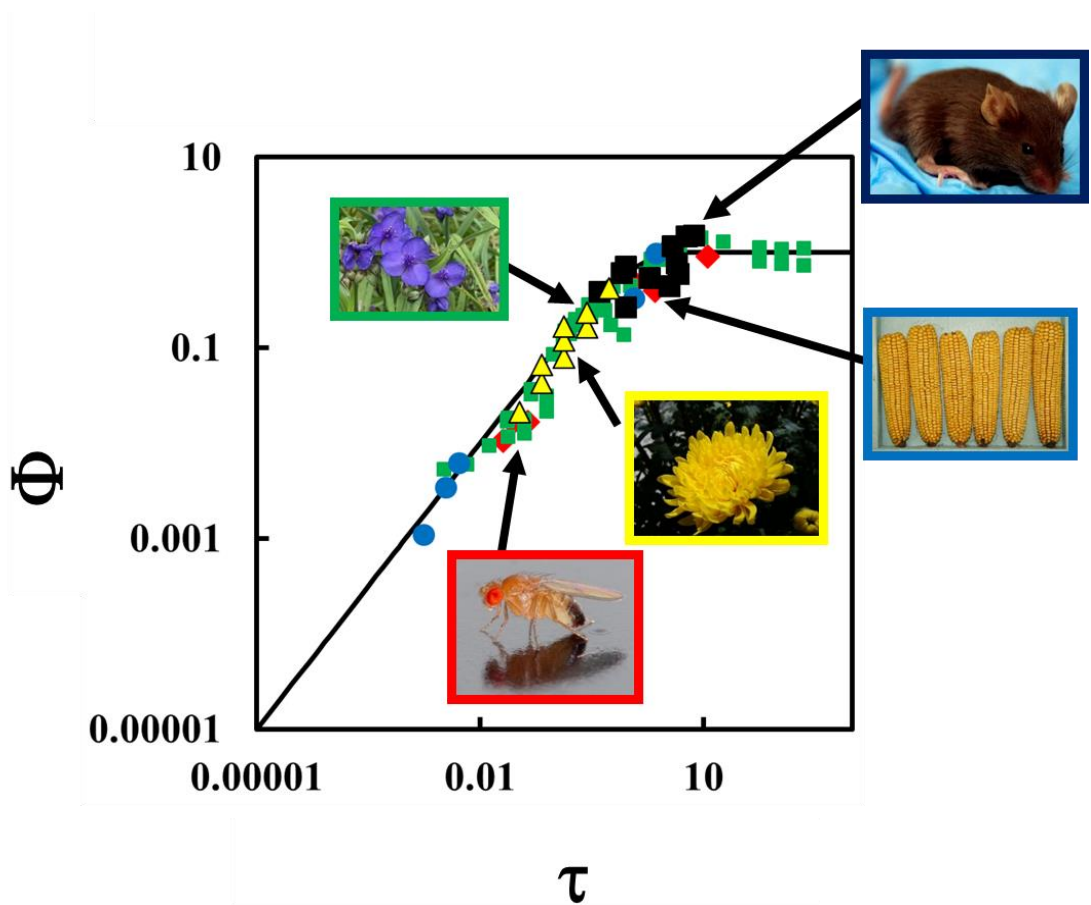

**Fig. 4. Comparison of experimental data with the theoretical curve in $(\tau, \Phi(\tau))$ plane.**